\documentclass[10pt, conference, compsocconf]{IEEEtran}
\pdfoutput=1

\usepackage{amsmath}
\usepackage{mathtools}
\usepackage{graphicx}
\usepackage{listings}
\usepackage{subfigure}
\usepackage{balance}
\usepackage{color}
\usepackage{verbatim}

\usepackage{url}
\usepackage[nocompress]{cite}
\usepackage{cite}

\begin{document}
%
\title{Bid-Centric Cloud Service Provisioning}

\author{\IEEEauthorblockN{
	Philip Healy\IEEEauthorrefmark{1},
	Stefan Meyer\IEEEauthorrefmark{1},
	John Morrison\IEEEauthorrefmark{1},
	Theo Lynn\IEEEauthorrefmark{2},
	Ashkan Paya\IEEEauthorrefmark{3},
	Dan C. Marinescu\IEEEauthorrefmark{3}} \\
	\IEEEauthorblockA{\IEEEauthorrefmark{1}Irish Centre for Cloud Computing \& Commerce\\ University College Cork, Ireland\\
	\{p.healy, s.meyer, j.morrison\}@cs.ucc.ie} \\
	\IEEEauthorblockA{\IEEEauthorrefmark{2}Irish Centre for Cloud Computing \& Commerce\\ Dublin City University, Ireland  \\ theo.lynn@dcu.ie} \\
	\IEEEauthorblockA{\IEEEauthorrefmark{3}University of Central Florida, USA\\
	ashkan\_paya@knights.ucf.edu, dcm@cs.ucf.edu}
}


\IEEEcompsoctitleabstractindextext{%
\begin{abstract}

Bid-centric service descriptions have the potential to offer a new cloud service provisioning model that promotes portability, diversity of choice and differentiation between providers. A bid matching model based on requirements and capabilities is presented that provides the basis for such an approach. In order to facilitate the bidding process, tenders should be specified as abstractly as possible so that the solution space is not needlessly restricted. To this end, we describe how partial TOSCA service descriptions allow for a range of diverse solutions to be proposed by multiple providers in response to tenders. Rather than adopting a lowest common denominator approach, true portability should allow for the relative strengths and differentiating features of cloud service providers to be applied to bids. With this in mind, we describe how TOSCA service descriptions could be augmented with additional information in order to facilitate heterogeneity in proposed solutions, such as the use of coprocessors and provider-specific services.

\end{abstract}

}

\maketitle

\IEEEdisplaynotcompsoctitleabstractindextext

%
\IEEEpeerreviewmaketitle

%

\section{Introduction}

Cloud service description languages are at an early stage of development and adoption. Nevertheless, technologies and standards are emerging that allow for abstract descriptions to be instantiated in a portable fashion on multiple cloud providers. This process occurs by mapping the service topology specified in the service description to the resource types, such as virtual machine specifications, available at the provider. The required resources are then provisioned and configured as necessary to bring the service online. However, the current state of the art is to adopt a lowest common denominator approach where only the generic features commonly supported by cloud providers are assumed.

In reality, the cloud providers' offerings are anything but generic, with competition spurring innovation and differentiation. Apart from their low-level Infrastructure-as-a-Service (IaaS) offerings, many providers offer additional services -- such as object stores, e-mail, messaging and monitoring -- that can be used in place of equivalent services running on virtual machines provisioned directly by the client. Higher-level services such as Amazon Elastic MapReduce can replace entire clusters of directly-provisioned virtual machines. Some providers allow for coprocessors, such as GPUs, to be attached to and used by virtual machines. This trend is expected to accelerate in the future, allowing for a variety of heterogeneous computing resources to be used by cloud services.

Rather than ignoring the diversity amongst cloud providers, it can be advantageous to embrace it. Using a provider's service for generic functionality, such as monitoring and database hosting, might be cheaper and more reliable than running a dedicated virtual machine. Other features, such as automated backups and scalability, might also make them attractive alternatives to providing the equivalent functionality directly as part of the service deployments. Compute-intensive services, such as genomics applications, may be able to take advantage of coprocessors if they are available.

The desire to maintain service portability through abstract service descriptions stands in conflict with the desire to take advantage of provider-specific services. We propose that this issue can be addressed by introducing a \textit{bid-centric} provisioning model. Under this approach, services are partially specified and submitted as a tender to multiple cloud providers, or to some brokers acting on their behalf. The partial description may then be fleshed out by each service provider based on the resources and services available, resulting in one or more bids. Each bid represents a complete service description that can be deployed with the corresponding provider when a winning bid is chosen. A sketch of our proposed bid-centric provisioning model is presented in Section \ref{sec:bid-centric-provisioning}. The bid-centric provisioning model, although independent, is motivated by the concept of self-organizing clouds, where autonomous resources configure themselves to form responses to bids. A generalized bid matching model is presented in Section \ref{sec:bid-matching-model} that can be applied to both scenarios.

In order to make the bidding model a reality, some method of describing tenders and bids is required. We argue that the existing TOSCA service description standard can be used to fulfill both roles. Section \ref{sec:service-description-languages} gives a brief overview of the current state of the art for service description languages. In Section \ref{sec:bid-centric-service-descriptions} we outline how TOSCA service descriptions can be used as the basis for a bid-centric provisioning model. Section \ref{sec:partial-descriptions} describes how partial TOSCA service descriptions can be used to implement a bidding process that promotes diversity amongst the resulting bids. Section \ref{sec:extended-descriptions} examines how the flexible tagging mechanism included in the TOSCA XML schema can be leveraged to include additional information that would allow for the bidding process to be opened up to include heterogeneous resources as coprocessors. Some simulation results based on our proposed bidding model are presented in Section \ref{sec:simulation_results}. Related work is examined in Section \ref{sec:related-work}. Finally, our conclusions and future work are presented in Section \ref{sec:conclusion}.


\section{Bid-centric Cloud Provisioning Model}

\label{sec:bid-centric-provisioning}

Traditionally, cloud resources have been provisioned using a direct purchasing model. Under the simplest scenario, customers purchase directly from a provider that they trust. This approach has the advantage of a rapid turnaround time and minimal evaluation effort, at the cost of ignoring potentially better offerings from rival providers. A more structured and involved evaluation process often begins with a research phase that gathers information on various providers' offerings through analysis of technical specifications, SLAs and pricing plans. The requirements of the service under consideration are then mapped to the offerings of the various providers, resulting in a number of potential instantiations of the service. A scoring mechanism is then used to decide which potential instantiation is best, using either a single parameter such as cost, or a more nuanced method such as a comparison table. The highest scoring instantiation may then purchased from the corresponding service provider. This approach places a burden on the customer, who must gather the information required to make an informed decision, then manually carry out the evaluation process.

In contrast, a bid-centric model would allow for the information gathering processing and the specification of potential instantiations to be performed externally, allowing the customer to focus on the service specification and the evaluation of the potential instantiations. This is achieved by following the tendering process as described in the Procurement literature \cite{dimitri2006handbook}:

\begin{enumerate}
\item The \textit{procurer} drafts a \textit{tender} describing a \textit{lot} and disseminates it to \textit{suppliers}.
\item Each supplier returns zero or more \textit{bids} in response to the tender.
\item The procurer evaluates the bids and chooses a \textit{winner} if all necessary conditions are met.
\item If a winning bid is selected, it is purchased from the corresponding supplier.
\end{enumerate}

In a Cloud Computing context, the roles of procurer and supplier map directly to the concepts of cloud consumer and cloud provider as defined by NIST \cite{mell2011nist}, and the lots being tendered are cloud service instantiations. A suitable mechanism is required for for describing services as tenders. Additionally, a corresponding mechanism must be found for describing bids in unambiguous terms so that they can be evaluated objectively. Rather than developing new service specification formats for this purpose, we argue that the existing TOSCA format can represent both tenders and bids if used appropriately. We note that other specification languages exist but few, if any, offer the advantages of TOSCA for capturing the bid-centric model.

Once suitable representation formats have been found for tenders and bids, the issue arises of how to conduct the procurement process. Ideally, the consumer would contact potential providers directly in order to notify them of the tender using a web service or some other suitable business-to-business communication method. There are two obvious drawbacks to this approach: it leads to a chicken-and-egg problem in terms of adoption, and the tender will only reach providers that the consumer is aware of. Alternatively, a brokerage service could be created that maintains a list of providers and contains sufficient knowledge about them to create bids on their behalf. This approach addresses the adoption issue but necessitates the creation of the brokerage service itself, and requires that the brokerage service maintains up-to-date information about the various provider offerings.

The broker concept could be extended to embody both of the previous approaches by forwarding tenders to providers that accept them but constructing bids itself on behalf of those that don't. This hybrid approach has the advantage of centralizing knowledge about the number of available providers, and their capabilities, while still allowing for providers to use internal information to construct bids. At the simplest level, this internal information could include up-to-date pricing and utilization data. However, more exotic schemes are possible, in particular the use of self-organization techniques to respond to individual tenders~\cite{marinescu2013auction}. We see this as an important future direction because it creates a new delivery model for Cloud Computing that augments the traditional IaaS, PaaS, and SaaS models.

After all the bids for a tender have been assembled, some mechanism for choosing a winner is required. As noted above, the evaluation process can become involved if multiple parameters are under consideration. A simple solution would be to present the customer with a table summarizing the values of parameters of interest for each bid. This approach has the benefit of simplicity but relies on human judgment, which may not be appropriate in all cases. An alternative would be to provide a decision support mechanism that uses a more formal process or, in the extreme case, for the system to automatically accept a bid if it meets a threshold of objectively specified criteria. As a starting point, a large body of related work is available on the use of structured decision making processes to evaluate bids \cite{de2001review}.

\section{Bid Matching Model}
\label{sec:bid-matching-model}

Next, we present a generalized capability-based bid matching model that can be used by suppliers to decide what bids, if any, should be submitted in response to tenders. Without loss of generality, this model can be used to implement the bid-centric cloud provisioning model described above. It could also form the basis of a self-organization model for cloud resources in response to service requests. The requirements of procurers and capabilities of suppliers are described by $m$ attributes, $\mathcal{A}_{1}, \mathcal{A}_{2},\ldots,\mathcal{A}_{i}, \ldots,\mathcal{A}_{m}$, where each attribute has at most $p$ possible realizations
\begin{equation}
\mathcal{A}_{i} = (a_{i,1}, a_{i,2}, \ldots, a_{i,j}, \ldots, a_{i,p}).
\end{equation}
In a Cloud Computing context, examples of attributes might include: system architecture with $32$- and $64$-bit realizations; the presence of CUDA-compatible GPUs, with the realizations being individual GPU card models; or, TOSCA node types, with the realizations being TOSCA node type implementations. Supplier $\mathcal{S}^{k}$ is described by an $m \times p$ {\it capability} matrix
\begin{align}
&C^{k} = \left(
\begin{array} {c}
c_{1}^{k} \\
\vdots \\
c_{i}^{k} \\
\vdots \\
c_{m}^{k}
\end{array}
\right)
~~c_{i}^{k} = (c_{i,1}^{k},\ldots, c_{i,j}^{k}, \ldots c_{i,p}^{k}) \nonumber \\[5pt]
&\text{and}~c_{i,j}^{k} = \left\{
\begin{array} {ll}
1 & \text{if~} a_{ij} \text{~ is supported for~} \mathcal{A}_{i} \\
0 & \text{otherwise}.
\end{array}
\right.
\end{align}
A tender $\mathcal{T}^{n}$ is described by an $m \times p$ {\it requirement} matrix
\begin{align}
&T^{n} = \left(
\begin{array} {c}
r_{1}^{n} \\
\vdots \\
r_{i}^{n} \\
\vdots \\
r_{m}^{n}
\end{array}
\right)
~~r_{i}^{n} = (r_{i,1}^{n},\ldots, r_{i,j}^{n}, \ldots r_{i,p}^{n}) \nonumber \\[5pt]
&\text{and}~ r_{i,j}^{n} = \left\{
\begin{array} {rl}
1 & \text{if~} r_{ij} \text{~ is desired for~} \mathcal{A}_{i} \\
0 & \text{if~} r_{ij} \text{~ is not desired for~} \mathcal{A}_{i} \\
-1 & \text{if~} r_{ij} \text{~ is not specified for~} \mathcal{A}_{i}.
\end{array}
\right.
\end{align}
The matching vector for supplier $\mathcal{S}^{k}$ and tender $\mathcal{T}^{n}$ is
\begin{equation}
\mu^{k,n} = \left( \mu_{1}^{k,n},\ldots,\mu_{i}^{k,n}, \dots, \mu_{m}^{k,n}   \right)~\text{with}~ \mu_{i}^{k,n}= c_{i}^{k} \cdot r_{i}^{n}.
\end{equation}
The matching vector $\mu^{k,n}$ can be used by supplier $\mathcal{S}^{k}$  to decide if it should submit a bid for tender $\mathcal{T}^{n}$. The number of nonzero components of $\mu^{k,n}$ indicates the number of attributes that $\mathcal{S}^{k}$ can satisfy and the number of zero components indicates the number of attributes it cannot satisfy.

\subsection{Example}
Consider the case when $m=3$; we have three attributes $\mathcal{A}_{1},\mathcal{A}_{2},\mathcal{A}_{3}$.  There are at most four realizations of each attribute, $p=4$. We assume that $\mathcal{S}^{k}$ offers: $a_{1,1}$ and $a_{1,2}$ for $\mathcal{A}_{1}$; $a_{2,3}$ for $\mathcal{A}_{2}$; and $a_{3,1}$ and $a_{3,3}$ for $\mathcal{A}_{3}$; the tender $\mathcal{T}^{n}$ demands: $a_{1,1}$ or $a_{1,2}$ for $\mathcal{A}_{1}$; $a_{2,1}$ or $a_{2,4}$ for $\mathcal{A}_{2}$; and it is satisfied with any choice for $\mathcal{A}_{3}$.  Then

\begin{align}
&C^{k} = \left(
\begin{array} {cccc}
1 & 1 & 0 & 0 \\
0 & 0 & 1 & 0 \\
1 & 0 & 1 & 0
\end{array}
\right) \nonumber \\
&\text{and} \nonumber \\
&T^{n} = \left(
\begin{array} {rrrr}
1 & 1 & 0 & 0 \\
1 & 0 & 0 & 1 \\
-1 & -1 & -1 & -1
\end{array}
\right) \nonumber
\end{align}
thus, the matching vector is
\begin{equation}
M^{k,n} = (2 ~~ 0 ~ -2). \nonumber
\end{equation}
Supplier $\mathcal{S}^{k}$ is able to provide two desired realizations for attributes $\mathcal{A}_{1}$ and substitute two choices for $\mathcal{A}_{3}$; it cannot satisfy $\mathcal{A}_{2}$. Therefore, no bids should be placed in this instance.

\subsection{Bid Quantification}

In order to estimate the size of the potential solution space for a tender, it is useful to quantify various attributes of the bid matching process. The number of unspecified attributes of tender $\mathcal{T}^{n}$ is
\begin{equation}
\nu^{n} = { 1 \over 2p}  \sum_{i=1}^{m} \sum_{j=1}^p (\mid r_{i,j}^{n} \mid - r_{i,j}^{n}).
\end{equation}
\noindent Given $N$ suppliers the probability that a randomly selected supplier $\mathcal{S}^{k}$ supports realization $a_{ij}$ for attribute $\mathcal{A}_{i}$ is
\begin{equation}
\label{ServerProb}
P_{i,j}^{k} = {\kappa_{i,j}^{k}  \over N} ~~\text~~\kappa_{i,j}^{k} = \sum_{k=1}^{N} c_{i,j}^{k},
\end{equation}
where $\kappa_{i,j}$ is the number of suppliers supporting realization $a_{ij}$ for attribute $\mathcal{A}_{i}$. The probability that a random tender $\mathcal{T}^{n}$  in a series of $M$ tenders is satisfied with the realization $a_{i,j}$ for attribute $\mathcal{A}_{i}$ is
\begin{equation}
\label{RequestProb}
Q_{i,j}^{n} = {1 \over M}\sum_{n=1}^{M} \mid r_{i,j}^{n} \mid
\end{equation}
if we assume that $\mathcal{T}^{n}$ can accept  any realization of the attribute $\mathcal{A}_{i}$ if it does not either specify or reject the realization $a_{i,j}$ of $\mathcal{A}_{i}$. The probability that a random tender $\mathcal{T}^{n}$ in a series of $M$ tenders is satisfied by a randomly selected supplier $\mathcal{S}^{k}$ in a set of $N$ suppliers is

\begin{equation}
V^{M,N} = \prod_{i=1}^{m} \prod_{j=1}^{p} P_{i,j}^{k} Q_{i,j}^{n}\text{.}
\end{equation}

\section{Cloud Service Description Languages}
\label{sec:service-description-languages}

Several service description languages (SDLs) are available that allow various aspects of services to be described. Some are specific to Cloud Computing, while otehrs are more generalized. A survey by Sun \textit{et al.}~found that the available SDLs exhibit wide variety in the level of abstraction used for service descriptions, semantic expressibility and their service coverage~\cite{sun_survey_2012}. Most languages lack support for the different cloud deployment models (such as public, private and hybrid), each of which comes with its own requirements, restrictions and dependencies. Each language tends to specialise in a particular area, such as operational, business or technical. In general, they lack a comprehensive specification model that covers disparate cloud resources, specific interaction interfaces and the actors involved in the service lifecycle (such as the service owners, consumers, and providers). We have found that the most comprehensive coverage of all aspects of cloud service deployments can be captured by using TOSCA and USDL in tandem. Together, they can capture the details of both the operational and business sides of a service, respectively.

\subsection{TOSCA}

TOSCA (Topology and Orchestration Specification for Cloud Applications) \cite{binz2012portable} is a cloud service standard by the Organization for the Advancement of Structured Information Standards (OASIS). TOSCA allows a service to be specified by an XML document that describes its topology, its components, and their relationships. Graphical editors are available that assist with the creation of description documents \cite{kopp2013winery}. Service components are modelled as nodes, which can represent software components as well as physical or virtual machines. Multiple nodes are typically used to describe a host and the software stack running on it. A key feature of the TOSCA model is that each node has one or more \textit{capabilities} and one of more \textit{requirements}. Nodes are connected together my matching requirements to capabilities, allowing for a loosely coupled design-by-contract model where nodes can be replaced by others with the same capabilities. The relationship between the nodes always has a source and a target and can have optional relationship constraints, such as \texttt{depends on} and \texttt{hosted on}. Multiple hosts can be abstracted into \textit{tiers} to facilitate horizontal scalability and failover mechanisms.

TOSCA allows for service lifecycle management to be specified by providing functionality for service creation, updates and termination. These so-called \textit{plans} are defined as process models, i.e., a workflow of one ore more steps. TOSCA relies on the established BPMN and BPEL standards to describe these workflows, but other languages can be used if desired. The implementation of the components of a plan is not fixed and can include shell scripts or configuration management scripts for tools such as Puppet or Chef. It is also possible to specify non-functional behaviour, such as SLAs. In TOSCA terminology, these are referred to \textit{policies}. Policies can also include monitoring behaviour, payment conditions, scalability or availability. The policies can be assigned to the whole service topology or individually to nodes. TOSCA applications are packaged as Cloud Service Archives (CSARs), which are ZIP files containing TOSCA definition files, plans and other resources required to deploy the service.


\subsection{TOSCA and USDL}

The Unified Service Description Language (USDL) \cite{cardoso2010towards} is a service description standard developed as a collaboration between several international research institutes, each of which contributed expertise from different backgrounds, such as Business, SLAs, Security and Computer Science~\cite{oberle_unified_2013}. The chief design objectives were conceptualization and modularity. The various service aspects are organized into packages, each of which represents a USDL module. Together, these modules model the business, operation and technical aspects of services.

Cardoso \textit{et al.}~evaluated the extent to which USDL and TOSCA could be combined to link the description and management of cloud services~\cite{cardoso_cloud_2013}. Whereas USDL can be used to enhance the service description for service discovery and selection, TOSCA helps service providers to automate the deployment and management of services. This allows for the creation of service marketplaces where service templates can be purchased and deployed to any one of a number of service providers. The combination of the two standards also allows for partial automation of the lifecycle management, such as discovery, selection, deployment and management. Under this scenario, different service level objectives can be specified for service templates, with the service topology specified using TOSCA and the SLA defined using USDL. Services specified in this fashion can cover a wide range of use cases and service types without compromising interoperability, portability or reversibility.

\section{Bid-centric Service Descriptions}
\label{sec:bid-centric-service-descriptions}

\begin{figure}[b]
\includegraphics[width=0.95\linewidth]{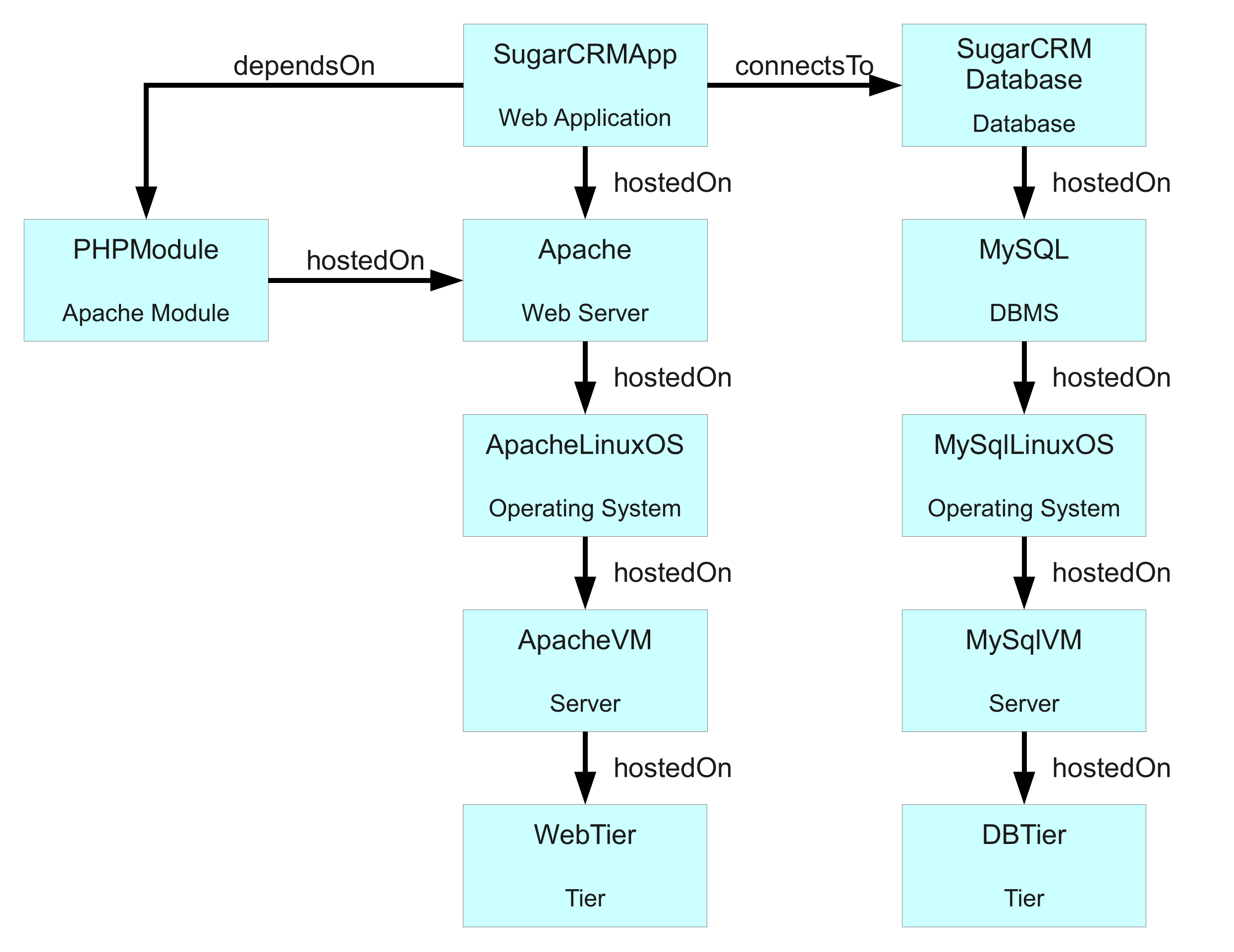}
\caption{Full two tier service architecture for a SugarCRM deployment as described in the TOSCA Primer.}
\label{fig:service-description-complete}
\end{figure}

As noted above, TOSCA can be used to completely specify the resources required to deploy a cloud service,  from application software at the top level down to operating systems and virtualized resources. Consider the two tier service description outlined in the TOSCA Primer \cite{oasis2013toscaprimer} depicted in Figure~\ref{fig:service-description-complete}. In the web tier, the application, web server, operating system and virtual machine are specified. Similarly, in the database tier the schema, database management system, operating system and virtual machine are specified. This completeness of description allows for services to deployed in a portable fashion across multiple cloud providers.

However, the portability afforded by complete service descriptions comes at a cost: by fully specifying all aspects of the service deployment, the diversity of potential solutions is curtailed.  Individual providers could differentiate in a number of ways, given the opportunity to do so:

\begin{itemize}
\item A provider might have PaaS offerings that could be substituted for some service requirements, such as web hosting and database management. For example, Amazon's Relational Database Service (RDS) could be substituted for the database virtual machine specified in the service description in Figure~\ref{fig:service-description-partial}. Substituting a provider-specific offering might lead to improved cost, availability and scalability.
\item A provider might have heterogeneous computing resources, such as GPUs, MICs and FPGAs, available as coprocessors. Application code may be able to take advantage of these if they are available.
\item A provider may offer additional functionality, such as detailed monitoring or backup services. It may be desirable to avail of this additional functionality if it is available.
\end{itemize}

In general, the portability of complete service descriptions is achieved by targeting the lowest common denominator in terms of functionality across service providers. In a bidding scenario, this prevents providers from differentiating based on their individual offerings. Furthermore, curtailing the solution space places limits on the novelty and diversity of the resulting bids. In contrast, a bid-oriented approach to describe services should strive to be as minimal as possible in order to allow for a wide range of potential solutions, and hence potential bids. Using the TOSCA service model, this ``opening up'' of the solution space can be achieved by partially specifying service descriptions so that they contain the minimum information required for a successful deployment, allowing for bids that fill in the blanks. This approach is described in Section~\ref{sec:partial-descriptions}. The solution space can be opened up further by enhancing service descriptions with additional information that provides hints to bidders about the range of acceptable solutions. In Section \ref{sec:extended-descriptions} we examine how this can be accommodated within the TOSCA specification.

\section{Partial Service Descriptions as Tenders}
\label{sec:partial-descriptions}

The TOSCA service description model centres around tiers, each of which is composed of nodes that have requirements and implement capabilities. Nodes represent composable service components. The top-level node of each tier typically has requirements. Requirements are met by adding nodes that provide the corresponding capability. However, nodes providing capabilities may in turn introduce transitive requirements. Dependencies may also exist between nodes in different tiers. As such, the process of creating a complete service description involves starting with one or more top-level nodes and iteratively adding nodes and tiers until all requirements are matched with corresponding capabilities.

TOSCA's requirements and capabilities model is powerful as it allows for a clean separation between service components, promoting modularity. The dependency relationship acts as a contract, allowing nodes to be substituted with differing implementations provided that required functionality is provided. In a bid-oriented system it is desirable to open the potential solution space to as many competing bids as possible. On the other hand, it is imperative that the resulting bids are capable of running the desired service. This can be achieved by using the TOSCA model to specify the minimal set of capabilities required to successfully deploy the service, while leaving non-essential capabilities unspecified to promote diversity amongst bids. The resulting bids would then fill out the unspecified functionality, resulting in a diverse collection of bids with differing approaches. However, the service descriptions in individual bids would be complete, allowing the service to be deployed when the winning bid is chosen.

From a bid matching perspective, each partially specified TOSCA node type can be modelled as an attribute $\mathcal{A}_{i}$ as described in Section \ref{sec:bid-matching-model}. The rows of the requirement matrix $\mathcal{T}^{n}$ for a tender are defined by the set of partially specified nodes in the service description. The realizations $a_{i,1},\ldots,a_{i,p}$ of each attribute are the set of possible implementations across all cloud providers. The capability matrix $\mathcal{C}^{k}$ for each provider is defined by the implementations supported for the different node types.

\begin{figure}
\includegraphics[width=0.95\linewidth]{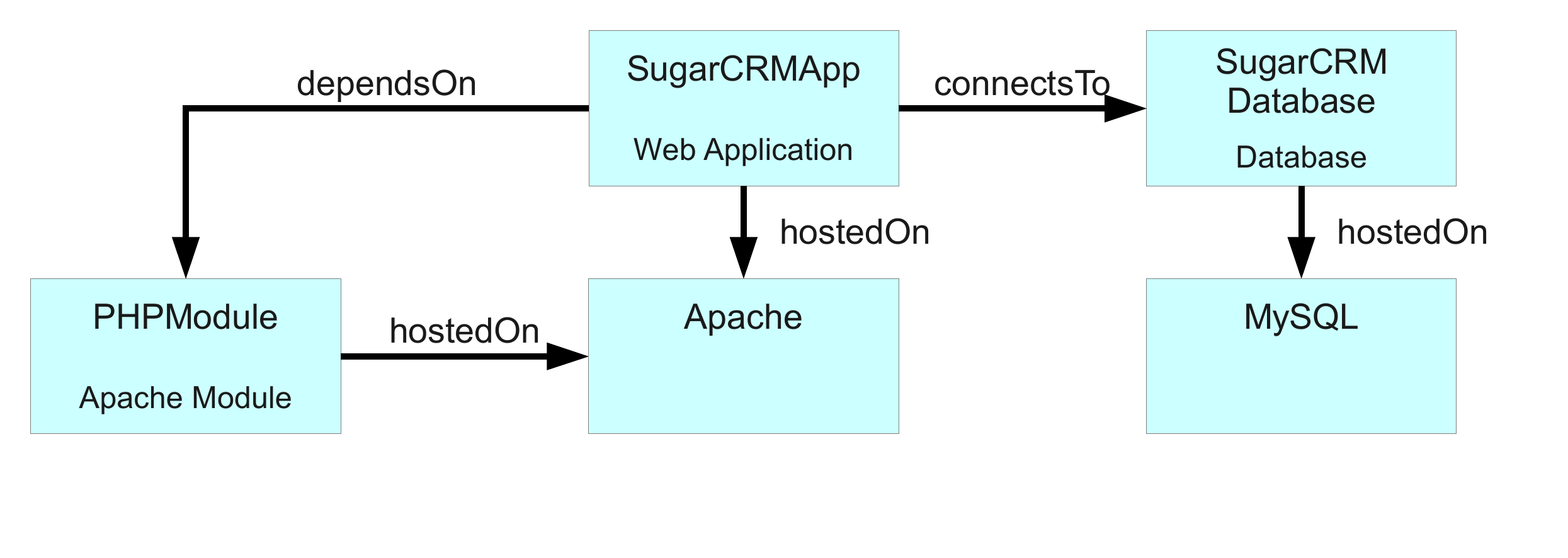}
\caption{A partial description of a SugarCRM service fully described in Figure \ref{fig:service-description-complete}.}
\label{fig:service-description-partial}
\end{figure}

\begin{figure}
\includegraphics[width=0.95\linewidth]{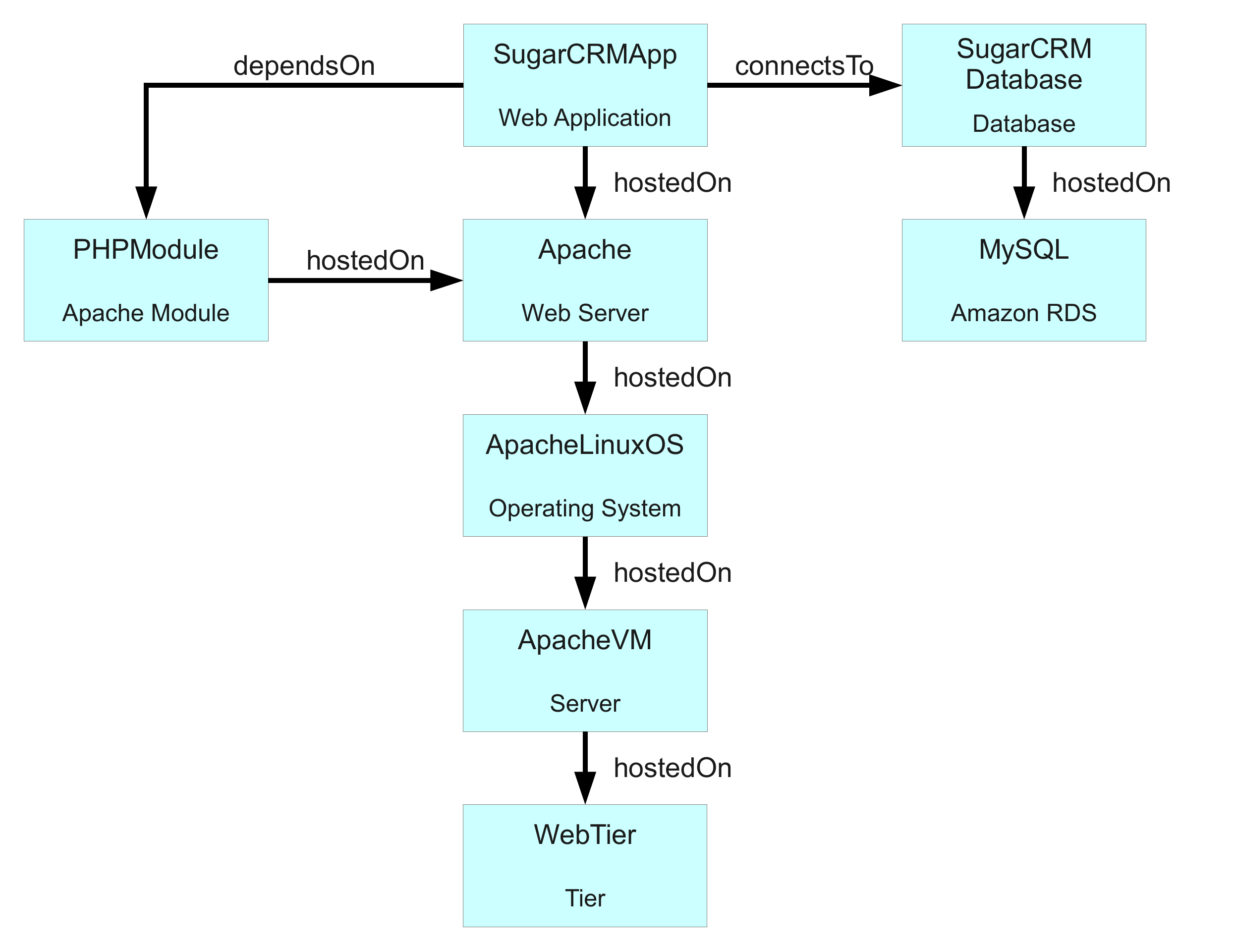}
\caption{An alternative instantiation of the SugarCRM service partially described in Figure \ref{fig:service-description-partial} that uses the Amazon Relational Database Service in lieu of a tier of virtual machines running the MySQL DBMS.}
\label{fig:service-description-alternate}
\end{figure}

Figure~\ref{fig:service-description-partial} depicts a partial description of the service fully described in Figure~\ref{fig:service-description-complete}. The building blocks that are necessary for a successful deployment are specified: the web application to deploy, the web server and PHP module that it depends on, and the database containing the required schema. Lower-level details, such as operating systems and virtual machines, are left unspecified. This partial service description would be submitted as part of a tender, and providers would return bids that satisfy the unspecified dependencies in various ways. If a provider has web hosting and/or database PaaS offerings, then these could be used instead of implementations based on virtual machines. Multiple bids based on virtual machines could be returned, such as a minimal cost implementation using a single VM and a high availability configuration with load balancing and redundancy. Figure~\ref{fig:service-description-alternate} depicts a bid for the tender specified in Figure~\ref{fig:service-description-partial} where integration with the Amazon Relational Database Service is used place of a tier of virtual machines running the MySQL DBMS.

\section{Extended Service Descriptions for Heterogeneous Resources}
\label{sec:extended-descriptions}

Some software packages, such as the SOAP genomics read alignment tool~\cite{liu2012soap3} and the Cycles 3D rendering engine~\cite{valenza2013blender}, can make use of GPUs for acceleration if available. Access to GPUs is already available from some commercial cloud providers, such as Amazon, with support also available in the OpenStack platform~\cite{crago2011heterogeneous}. As many integrated core (MIC) architectures, such as Intel’s Xeon Phi, become more widespread, it has been predicted that software will be increasingly adapted to make use of these resources given the lower development effort required compared to GPU acceleration~\cite{adie2012accelerators}. Although not widely supported at present, FPGAs are another coprocessor type that could be made available to cloud consumers~\cite{madhavapeddy2011reconfigurable}, with technical advances such as partial reconfiguration being proposed as enabling technologies~\cite{opitz2012accelerating}. More generally, it has  been predicted that coprocessing via heterogeneous computing resources will be increasingly supported by cloud providers in the future~\cite{singh2011computing}.

The non-uniformity of the coprocessing resources available at different cloud providers raises the question of how these can be incorporated into the bidding process. The existence of cloud service providers with no coprocessing resources available implies that a lowest common denominator approach cannot assume that any coprocessing resources are available. An alternative is to mandate the presence of coprocessors of a particular type, for example the NVIDIA Tesla M2050 GPUs currently offered by Amazon. However, this approach constrains the solution space -- it may be possible to run the application unaccelerated (e.g., the Cycles rendering engine) or it may be possible to accelerate using a number of coprocessor types, families or models. It is therefore necessary to provide a mechanism for augmenting the minimal description of a service with additional information that invites bids that include coprocessing resources. From a bid matching perspective using the model described in Section \ref{sec:bid-matching-model}, this additional information has the effect of increasing the number of potential realizations of each attribute $\mathcal{A}_{i}$.

A similar issue has arisen in the context of matching jobs to compute nodes in HPC clusters. The Condor distributed computing platform has a matchmaking system that matches jobs to machines capable of executing them~\cite{raman2003policy}. A classified advertisement (\textit{classad}) language is used to specify the characteristics, constraints and preferences of the entities to be matched. Classads are expressed as a collection of \mbox{name/value} pairs, with a sophisticated query-like syntax available for expressing constraint values. The classad matching scheme has been used to support GPU acceleration, allowing detailed information such as supported APIs, technical specifications and characteristics to be matched against\footnote{See the \textit{``How to Manage GPUs''} entry in the HTCondor Wiki.}. An example of a GPU matching classad is provided in Listing \ref{lst:gpu-classad}.

\begin{lstlisting}[caption=Condor classad excerpt indicating a requirement for a GPU with at least 16 cores that supports CUDA., label=lst:gpu-classad]
Requirements=HAS_GPU \
    && (GPU_API == "CUDA") \
    && (GPU_NUM_CORES >= 16)
\end{lstlisting}

The Condor classad scheme demonstrates that complex coprocessor capabilities can be expressed and matched against using a simple name/value pair specification scheme. The question now arises as to how such a scheme can integrated with TOSCA service descriptions.  Conveniently, the TOSCA specification includes a mechanism for associating arbitrary name/value pairs with the definitions of services, node types, and node type implementations (see Listing \ref{lst:name-value-schema}). By annotating node types definitions with tags describing supported coprocessors, detailed acceleration capabilities can be expressed while remaining within the confines of the TOSCA specification. Bidders could then take advantage of this information when responding to tenders.

\begin{lstlisting}[caption=Excerpt from the TOSCA XML schema definition.  Arbitrary name/value pairs can be associated with node types., label=lst:name-value-schema]
<NodeType name="xs:NCName"
          targetNamespace="xs:anyURI"?
          abstract="yes|no"? final="yes|no"?>
<Tags>
  <Tag name="xs:string" value="xs:string"/> +
</Tags> ?
\end{lstlisting}

Listing \ref{lst:name-value-example} provides an example of how such a scheme might work in practice. The presence of \texttt{gpu.support} tag with \texttt{optional} value indicates that this node type can be accelerated with a GPU if available, but that a GPU does not need to be present for the node type to function. The \texttt{gpu.cudaSupport} and \texttt{gpu.minCudaVersion} indicate that CUDA versions 3.0 and greater are supported. The \texttt{gpu.minNumCores} and \texttt{gpu.maxNumCores} tags indicate the number of GPU cores that can be usefully employed, indicating which models are most appropriate. The \texttt{gpu.multiCardSupport} indicates that the required number of cores can be made up by attaching multiple cards. Suitable tags could be included for specifying particular GPU families and models. Tags could also specify memory charecteristics, such as the amount of memory per card and the memory type (such as GDDR version and ECC support).

\begin{lstlisting}[caption=Proposed TOSCA node type tags for GPU acceleration, label=lst:name-value-example]
<Tags>
  <Tag name="gpu.support" value="optional"/>
  <Tag name="gpu.cudaSupport" value="true"/>
  <Tag name="gpu.minCudaVersion" value="3.0"/>
  <Tag name="gpu.minNumCores" value="448"/>
  <Tag name="gpu.maxNumCores" value="5000"/>
  <Tag name="gpu.multiCardSupport" value="true"/>
</Tags>
\end{lstlisting}

Similar tagging schemes could be put in place for other coprocessor types, such as MICs and FPGAs. The attributes for specifying MIC support would be similar to those for GPUs given the similarity in their characteristics. However, attributes for FPGAs would need to be more specific, given the fact that the place-and-route process is typically targeted to a particular FPGA model.

\section{Simulation Results}
\label{sec:simulation_results}

We now report on the results of simulation experiments based on the model introduced in Section \ref{sec:bid-centric-provisioning}. In our simulation a tender can be described by up to $m=50$ attributes and there are $p=10$ possible realization of each attribute. The $N=100$ suppliers provide $10^{5}$ bids for the $1,000$ tenders. We display the success ratio and group the tenders in bins of $2,000$ and the average number of unspecified attributes in bins of $20$ for several capability and requirement matrices.

\begin{figure*}[p]
\begin{center}
\includegraphics[width=7.5cm]{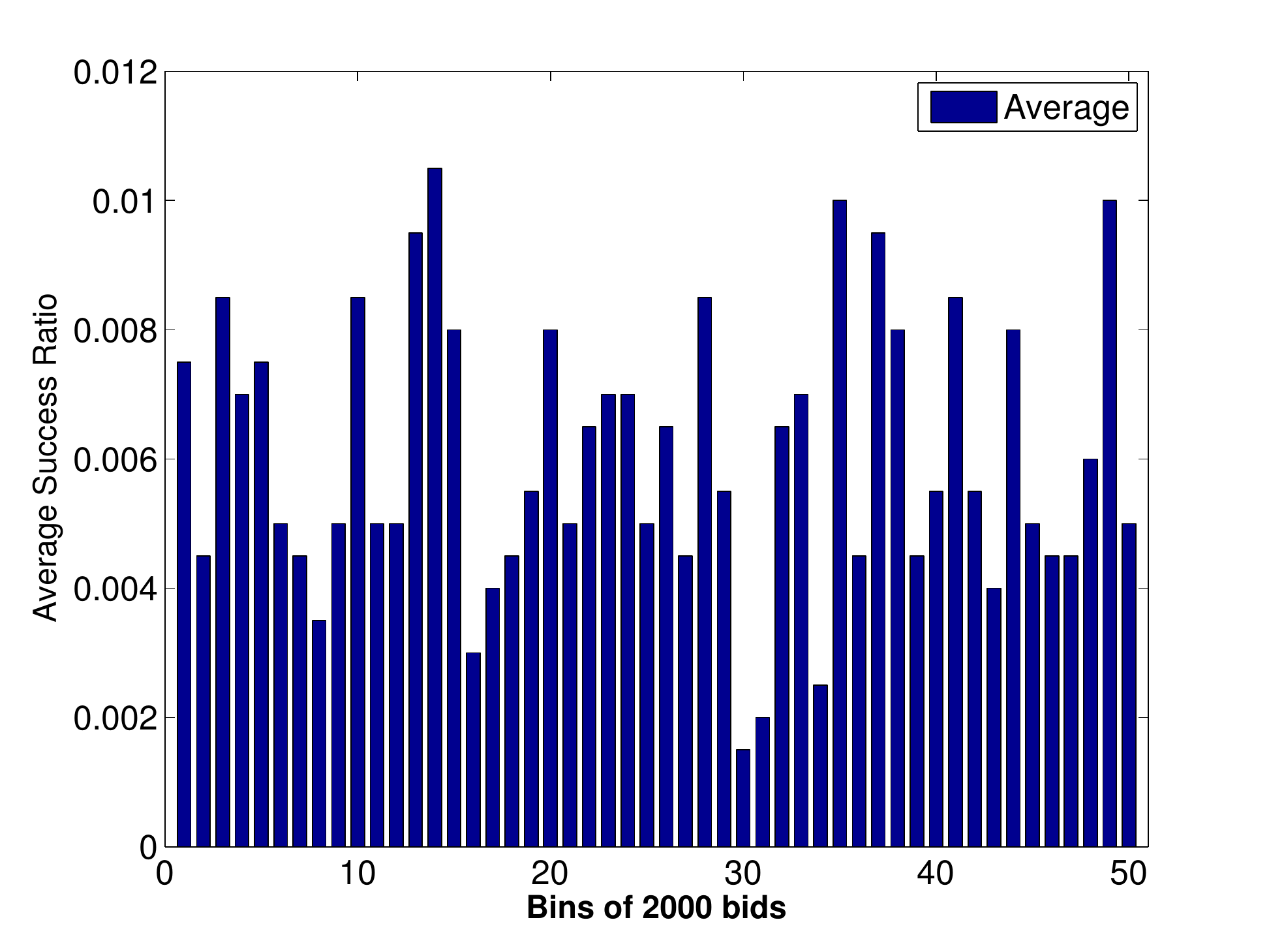}
\includegraphics[width=7.5cm]{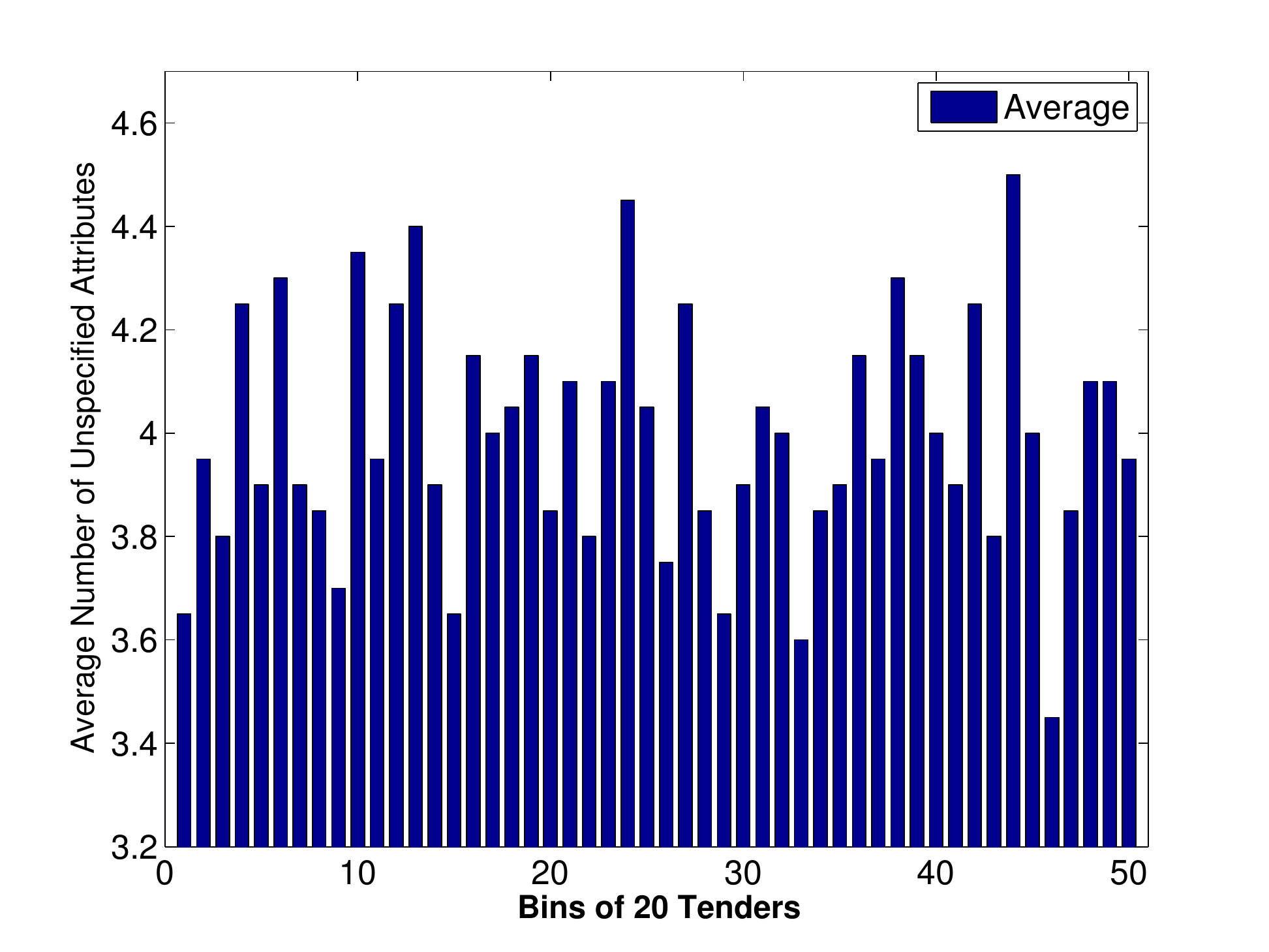}\\
~~~~~~~~~~~~~(a)~~~~~~~~~~~~~~~~~~~~~~~~~~~~~~~~~~~~~~~~~~~~~~~~~~~~~~~~~~~~(b)
\end{center}
\caption{$a_{i,j}^{k}$ uniformly distributed in [0.1 0.8]. (a) Average success ratio of tenders. (b) Average number of unspecified attributes for tenders.}
\label{0108}
\end{figure*}

\begin{figure*}[p]
\begin{center}
\includegraphics[width=7.5cm]{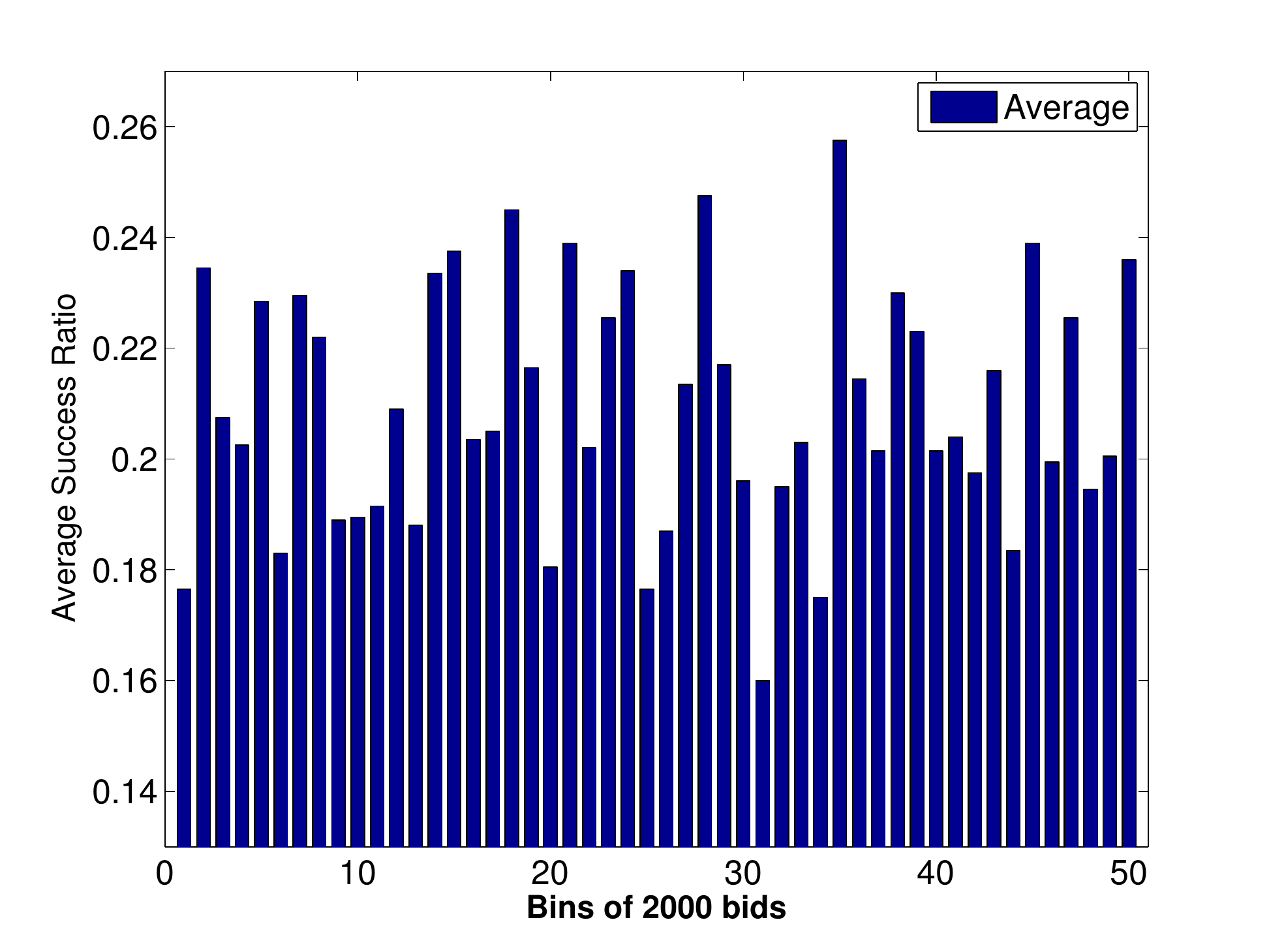}
\includegraphics[width=7.5cm]{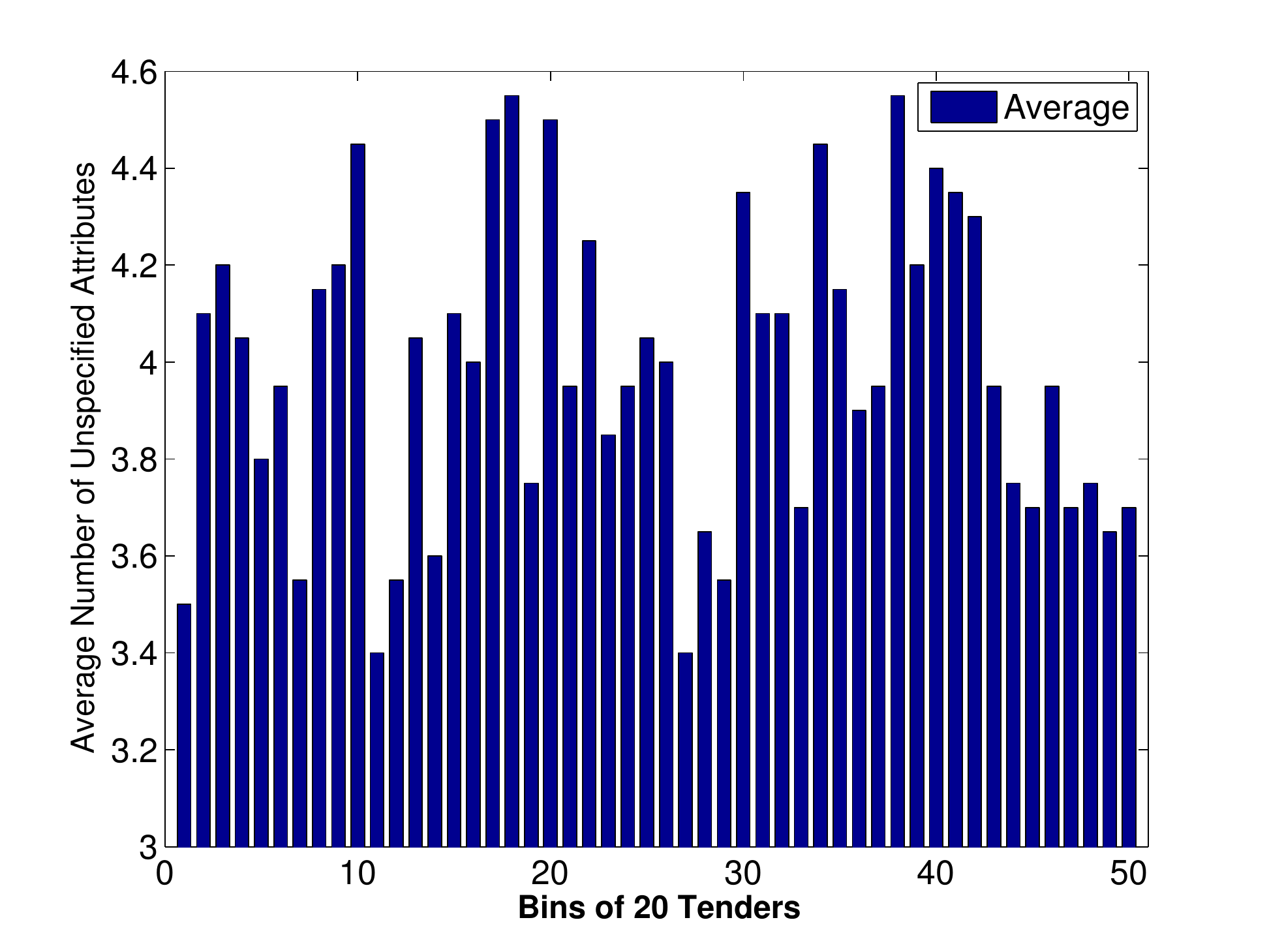}\\
~~~~~~~~~~~~~(a)~~~~~~~~~~~~~~~~~~~~~~~~~~~~~~~~~~~~~~~~~~~~~~~~~~~~~~~~~~~~~~~~~(b)
\end{center}
\caption{$a_{i,j}^{k}$ uniformly distributed in [0.3 0.8]. (a) Average Success rate of tenders. (b) Average number of unspecified attributes for tenders.}
\label{0308}
\end{figure*}

\begin{figure*}[p]
\begin{center}
\includegraphics[width=7.5cm]{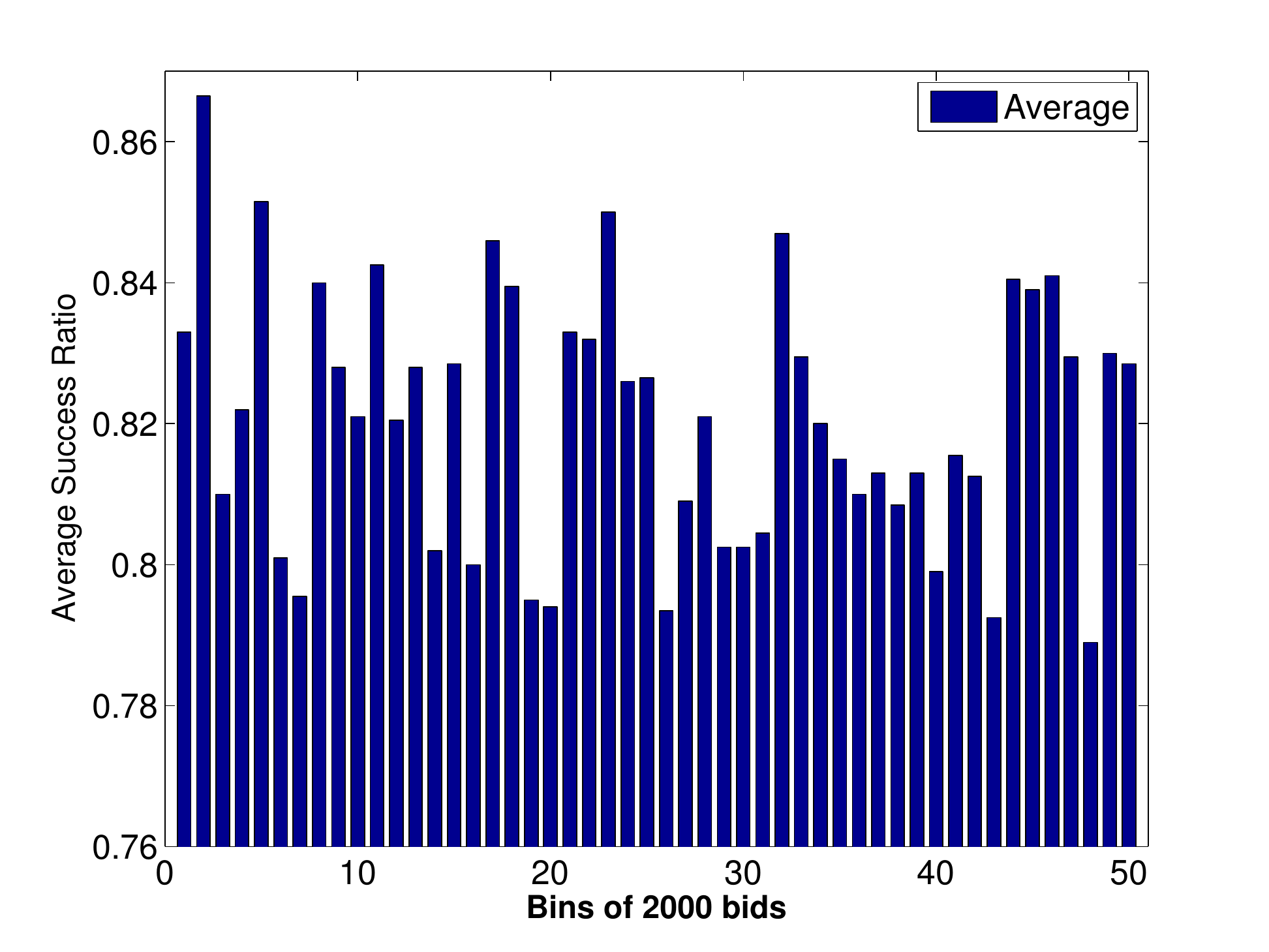}
\includegraphics[width=7.5cm]{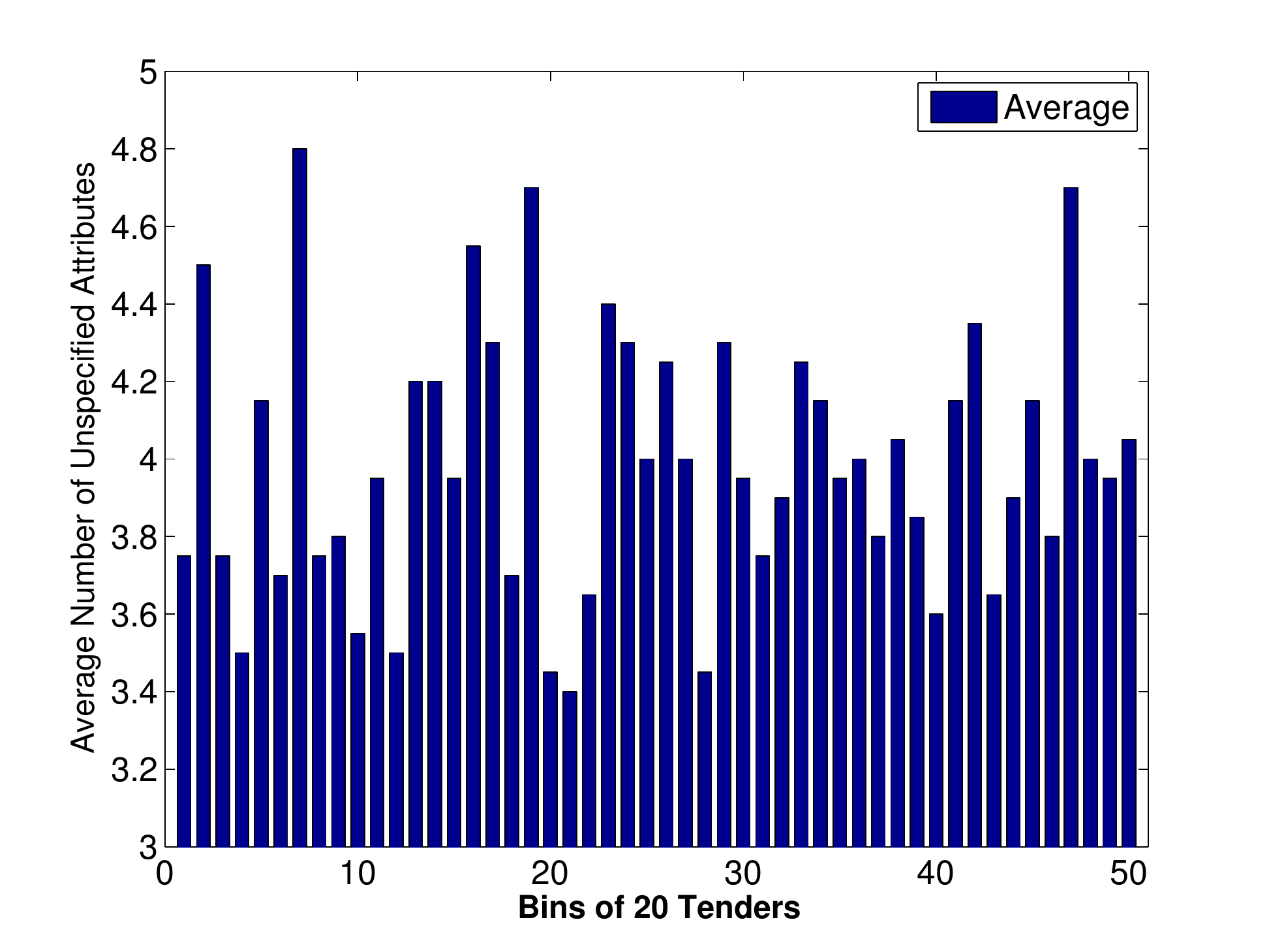}\\
~~~~~~~~~~~~~(a)~~~~~~~~~~~~~~~~~~~~~~~~~~~~~~~~~~~~~~~~~~~~~~~~~~~~~~~~~~~~~~~~~(b)
\end{center}
\caption{$a_{i,j}^{k}$ uniformly distributed in [0.5 0.8].(a) Average Success. (b) Average number of unspecified attributes for tenders.}
\label{0508}
\end{figure*}

In the first experiment, the probability that the supplier $S^k$ supports attribute $a_{i,j}^k$ for $C^{k}$ is uniformly distributed in the interval [0.1, 0.8]; the number of unspecified attributes of tender $T^{n}$ is uniformly distributed in the range of [2,6]. Figures~\ref{0108}(a) and (b) show the average success ratio and the average number of unspecified attributes displayed, respectively. In the next two experiments, see Figures~\ref{0308} and ~\ref{0508}, the attributes are uniformly distributed in the intervals [0.3, 0.8] and [0.5, 0.8], respectively.

These results show that in the first experiment the average success rate is very low, in the $2 - 10\%$ range due to the low average probability $\bar{p}(a_{i})=0.45$ that a server supports a given attribute. As this average increases to $0.55$ and then to $0.65$ the range of the success ratios are much higher, $[19 - 26\%$ and $79 - 83\%$, respectively. The distribution of the number of unspecified attributes seems to be invariant.

\section{Related Work}
\label{sec:related-work}

ABACUS \cite{zhang2013abacus} is a resource management framework that allows for cloud service differentiation based on job characteristics. Each job submission has an associated budget and utility function. The utility function is used to indicate the benefit accrued by allocating the job to sets of resources. When resources become available, these parameters are used to decide which outstanding job they will be allocated to. Experimental results based on a MapReduce use case are presented.

Shi \textit{et al.} present an electronic auction platform for cloud resources based on a continuous double auction mechanism \cite{shi2013continuous}. The platform uses trading rounds to match bids from consumers with asks from cloud service providers. A two stage game bidding strategy is also presented. Song \textit{et al.} present another market model based on combinatorial auctions \cite{song2009novel}. This model allows for collaboration between service providers when creating bids. Service providers can autonomously find partners and create groups that increase their competitive power and hence improve their chances of submitting a winning bid. 

The MODAClouds project \cite{ardagna2012modaclouds} seeks to develop a a model-driven approach for the design and execution of applications across multiple clouds. Under this approach, applications are developed at a high level that abstracts the capabilities of the clouds that may be targeted during deployment. These high-level specifications are then semi-automatically translated to run on multiple cloud platforms, allowing for flexibility in terms of cost, risk and quality of service.

\section{Conclusions and Future Work}
\label{sec:conclusion}

Rather than focusing simply on the functionality of the service descriptions, future work will consider the  inclusion of service level objectives, such as cost and response time, as bid criteria. A large body of existing work is available on the topic of machine readable SLAs and automated SLA negotiation~\cite{hasselmeyer2007, wieder2011service}. By combining existing machine-readable SLA technologies such as WS-Agreement and SLA*, it should be possible to incorporate SLA negotiation into the bidding process.

Many cloud services are not mapped to a static set of resources for the operational phase of their lifecycle -- the on-demand nature of cloud computing allows them to scale the resources used in response to demand. The issue of incorporating elasticity into TOSCA service descriptions and the TOSCA runtime have already been addressed by the ElasticTOSCA project~\cite{han2013elastic}. Future work will examine how elasticity can be incorporated into the tendering and bidding processes.

The question of bid evaluation by consumers will also be examined through the application of formal evaluation techniques such as the analytic hierarchy process \cite{saaty1988analytic}. This would allow for objective evaluation based on the consumer's particular preferences. This preference information could also be incorporated into the tendering process, allowing service providers to tailor their bids to the individual requirements of the consumers.

\section{Acknowledgement}
The research work described in this paper was supported by the Irish Centre for Cloud Computing and Commerce, an Irish national technology centre funded by Enterprise Ireland and the Irish Industrial Development Authority.

\ifCLASSOPTIONcaptionsoff
  \newpage
\fi


\balance
\bibliographystyle{IEEEtran}
\bibliography{IEEEabrv,Bibliography}

%






\end{document}